\documentclass{emulateapj}

\usepackage{url}

\usepackage{epsf}

\newcommand{\myemail}{Email: obenvenu@fcaglp.unlp.edu.ar}

\shorttitle{Single Degenerate Type Ia Supernovae}

\shortauthors{O. G. Benvenuto, et al.}

\begin{document}

\title{Final Evolution and Delayed Explosions of Spinning White Dwarfs\\
 in Single Degenerate Models for Type I\lowercase{a} Supernovae}
 \submitted{Accepted for publication in the Astrophysical Journal letters}

\author{Omar G. Benvenuto\altaffilmark{1} and Jorge A. Panei\altaffilmark{2}}

\affil{Facultad de Ciencias Astron\'omicas y Geof\'{\i}sicas,
  Universidad Nacional de La Plata and Instituto de Astrof\'{\i}sica
  de La Plata, CCT-CONICET-UNLP. Paseo del Bosque S/N (B1900FWA), La
  Plata, Argentina}

\author{Ken'ichi Nomoto\altaffilmark{3}} \affil{Kavli Institute for
  the Physics and Mathematics of the Universe (WPI), The University of
  Tokyo\\ Kashiwanoha 5-1-5, Kashiwa, Chiba 277-8583, Japan}
                   
\author{Hikaru Kitamura} \affil{Department of Physics, Kyoto
  University, Sakyo-ku, Kyoto 606-8502, Japan}

\and

\author{Izumi Hachisu} \affil{Department of Earth Science and
  Astronomy, College of Arts and Sciences, The University of Tokyo\\
  Komaba 3-8-1, Meguro-ku, Tokyo 153-8902, Japan}

\altaffiltext{1}{Member of the Carrera del Investigador
  Cient\'{\i}fico of the Comisi\'on de Investigaciones
  Cient\'{\i}ficas (CIC) de la Provincia de Buenos Aires,
  Argentina. \myemail}

\altaffiltext{2}{Member of the Carrera del Investigador
  Cient\'{\i}fico y Tecnol\'ogico, CONICET, Argentina.}

\altaffiltext{3}{Hamamatsu Professor}

%-------------------------------------------------------------------------------

\begin{abstract}
We study the occurrence of delayed SNe~Ia in the single
degenerate (SD) scenario. We assume that a massive carbon-oxygen
(CO) white dwarf (WD) accretes matter coming from a companion star,
making it to spin at the critical rate. We assume uniform rotation due
to magnetic field coupling. The carbon ignition mass for non-rotating WDs
is $M_{\rm ig}^{\rm NR} \approx
1.38\ M_{\odot}$; while for the case of uniformly rotating WDs it is a few percent larger ($M_{\rm ig}^{\rm R} \approx
1.43\ M_{\odot}$). When accretion rate decreases, the WD begins to
lose angular momentum, shrinks, and spins up; however, it does not
overflow its critical rotation rate, avoiding mass shedding. Thus,
angular momentum losses can lead the CO WD interior to compression
and carbon ignition, which would induce an SN~Ia. The
delay, largely due to the angular momentum losses timescale, may be 
large enough to allow the companion star to evolve to a He
WD, becoming undetectable at the moment of explosion.  This scenario
supports the occurrence of delayed SNe~Ia if the final CO WD mass is
$1.38\ M_{\odot} < M < 1.43\ M_{\odot}$.
We also find that if the delay is longer than $\sim 3$ Gyr, the WD
would become too cold to explode, rather undergoing collapse.
\end{abstract}


%-------------------------------------------------------------------------------
\section{INTRODUCTION} \label{sec:introd}

The origin of Type Ia supernovae (SNe~Ia) has been one of the central
issues of recent astrophysics.
The main features of SNe~Ia have been firmly accounted for
as a consequence of a thermonuclear explosion of a carbon-oxygen (C+O)
white dwarf (WD).  Both Chandrasekhar mass 
and sub-Chandrasekhar mass models
have been presented (\citealt{2000ARA&A..38..191H};
\citealt{2000AIPC..522...35N}).  However, 
it is still not clear if the the WD accretes
H/He-rich matter from its binary companion (single degenerate
(SD) scenario), or two C+O WDs merge (double degenerate (DD)
scenario) \citep{2014ARA&A..52..107M}.

Observations provided some constraints on the nature of companion
stars that may be considered as an indication of the occurrence of the SD
scenario. These are the presence of circumstellar matter (CSM)
\citep{2007Sci...317..924P, 2011Sci...333..856S, 2012ApJ...752..101F}
and the detection of hydrogen in the circumstellar-interaction-type
SNe~(Ia/IIn), such as SN~2002ic \citep{2003Natur.424..651H} and PTF11kx
\citep{2012Sci...337..942D}.  On the other hand, it has been not
possible to detect the presence of companions, e.g., the lack of
companion stars in the images of SN~2011fe
\citep{2011Natur.480..348L}, some SN~Ia remnants (SNRs)
\citep{2012Natur.481..164S}, SN 1572 (Tycho,
\citealt{2009ApJ...701.1665K}), and SN 1006 \citep{2012ApJ...759....7K}.

However, the above discussion does not take into account that the
accreting WD should be rotating by getting the angular momentum from
the accreting materials.  \citet{2011ApJ...730L..34J} and
\citet{2011ApJ...738L...1D} presented a spin-down scenario and
suggested that there should exist a long spin-down phase of the
rapidly rotating WD with a timescale of the angular momentum loss
($J$-loss) from the WD.  If so, the donor star in the SD model may
exhaust its hydrogen-rich envelope and become a He WD before the SN~Ia
explosion.  \citet{2012ApJ...756L...4H} and
\citet{2014MNRAS.445.2340W} studied close binary evolutions by taking
into account the rotation of accreting WDs and obtained the parameter
space (the companion's mass and the binary separation) where the
``$J$-loss'' evolution actually occurs. The structure and evolution of 
rotating WDs have been studied by many groups in 1D
\citep[e.g.,][]{1968ApJ...151.1089O, 2000A&A...362.1046L,
pie03, sai04, yoo04, Pir08} and 2D \citep[e.g.,][]{hac86,
2003ApJ...595.1094U, 2005A&A...435..967Y}.

In the present study, we have constructed a new 1D stellar evolution
code by taking into account rotation.  In
Section~\ref{sec:methods} we describe the ingredients of the models as
well as the numerical techniques employed.  With this code, we have
calculated the final evolution of the mass-accreting WDs by assuming
that it rotates uniformly because of magnetic field coupling.  By
combining the close binary scenario \citep{2012ApJ...756L...4H}, we
have found that the WD evolves as follows, as it will be described in
detail in Section~\ref{sec:resultados}.

\noindent
(1) For certain ranges of binary parameters, as it will be presented in
Section~\ref{sec:discu_conclu}, the accretion rate ($\dot{M}$) always exceeds
$10^{-7} M_{\odot}$ yr$^{-1}$ so that the WD increases its mass until
it undergoes ``prompt'' carbon ignition.  The mass of the uniformly
rotating WD at the carbon ignition, $M_{\rm Ig}^{\rm R}$, is larger
for smaller $\dot{M}$.  For $\dot{M} = 10^{-7}
M_{\odot}$ yr$^{-1}$, $M_{\rm Ig}^{\rm R} =$ 1.43~$M_{\odot}$, which is
the largest mass because nova-like hydrogen flashes prevent the the WD
mass from growing for the lower $\dot{M}$.  Because of the
centrifugal force in the rotating WD, $M_{\rm Ig}^{\rm R}=$
1.43~$M_{\odot}$ is larger than $M_{\rm Ig}^{\rm NR} =$
1.38~$M_{\odot}$ \citep{1984ApJ...286..644N}.

\noindent
(2) For adjacent ranges of binary parameters
(\S\ref{sec:discu_conclu}), the mass of the rotating WD exceeds
$M_{\rm Ig}^{\rm NR} =$ 1.38~$M_{\odot}$ but does not reach $M_{\rm
  Ig}^{\rm R}=$ 1.43~$M_{\odot}$ because of the decreasing accretion
rate.  In this Letter, we shall present the evolution of the WD
whose mass reaches 1.40~$M_{\odot}$, being slightly below $M_{\rm
  Ig}^{\rm R}$.

After the accretion rate falls off, the WD undergoes the $J$-loss
evolution.  The exact mechanism and the timescale of the $J$-loss
are highly uncertain, although we consider that the 
magneto-dipole braking WD to be responsible.  We have found that $J$-loss induces
the contraction of the WD, which leads to the ``delayed'' carbon
ignition after the ``delay'' time due to neutrino and radiative
cooling.  

We also studied how the carbon ignition depends on the strong
screening factor, which still involves some uncertainties.  Then, we
shall discuss whether the WD undergoes the SN Ia explosion or collapse
(\S\ref{sec:resultados}).  Finally, we discuss our
findings in Section~\ref{sec:discu_conclu}.

%-------------------------------------------------------------------------------
\section{METHODS} \label{sec:methods}

We have employed the hydrostatic stellar evolution code
described in \citet{2013ApJ...762...74B}, adapted to the case of
accreting WDs. The equation of state of degenerate electrons is
treated as in \citet{1964ZA.....60...19K}, while Coulomb corrections
at finite temperature are included following
\citet{1973PhRvA...8.3096H}. Radiative opacities are taken from
\citet{1996ApJ...464..943I} whereas conductive opacities were included
following \citet{2008ApJ...677..495I}. Neutrino emission processes are
those described in \citet{1996ApJS..102..411I}.

As stated above, in this Letter we consider uniform rotation. For each isobar, we computed
the parameter $\lambda$, defined as
\begin{equation}
 \lambda= \frac{\Omega^2 r_{P}^{3}}{2 G M_{P}},
\end{equation}
\noindent
where $\Omega$ is the angular rotation, $r_{P}$ is the radius of a
sphere with a volume equal to the corresponding isobar, and $M_{P}$ is
the embraced mass. $\lambda$ measures the fastness of rotation;
breakup rotation rate corresponds to $\lambda \approx 0.270$.

Let us define the average $<q>$ of an attribute $q$ on a given isobar as
\begin{equation}
 <q>= \bigg(\frac{1}{S_{P}}\bigg) \oint_{P} q\; d\sigma
\end{equation}
\noindent
where $S_{P}$ is the isobar surface and $d\sigma$ an
element of it.

In order to compute the structure of the rotating star, we need the
coefficients $f_{P}$ and $f_{T}$ that correct the standard structure
equations for non-rotating objects. These are
\begin{eqnarray}
 f_{P}= \bigg(\frac{4\pi r_{P}^{4}}{GM_{P}S_{P}}\bigg) \frac{1}{<g_{\rm eff}^{-1}>},
 \nonumber \\
 f_{T} = \bigg(\frac{4\pi r_{P}^{2}}{S_{P}}\bigg)^{2} 
          \frac{1}{<g_{\rm eff}^{-1}>\; <g_{\rm eff}>}
\end{eqnarray}
\noindent
where $g_{\rm eff}$ is the effective gravitational acceleration, computed
in the Roche model approximation \citep{2009pfer.book.....M}. 

When the WD is accreting, we impose the critical
rotation regime for $\lambda= 0.270$ for the outermost layers. This
provides the (uniform) rotation rate of the entire star.

For the mass accretion rate, we assume that $\dot{M}$ is constant
during the main accretion stage and then declines exponentially with
an assumed timescale $\tau_{\dot{M}}=4\times10^{4}$~years (set in order to
mimic binary evolution) since a moment chosen in order for the WD
to reach an assumed final mass value $M$.  Since $\dot{M}\leq 1\times
10^{-7}\; M_{\odot}$ yr$^{-1}$, we also consider
$J$-losses with another timescale $\tau_{J}$, so that
\begin{equation}
 J= J(t_{0})\exp{(-(t-t_{0})/\tau_{J})},
\end{equation}
\noindent
where $t_{0}$ is the age at which $\dot{M}= 1\times 10^{-7}\;
M_{\odot}$ yr$^{-1}$. We have chosen this prescription to explore the
reaction of the WD to $J$-losses without assuming a particular
physical process, which at present is uncertain. For this
reason, we shall treat $\tau_{J}$ as a free parameter.  A possible
mechanism leading to such losses is magneto-dipole braking
(\citealt{2012MNRAS.419.1695I}; \citealt{2013ApJ...772L..24R}).
For $t>t_{0}$ we compute the rotation rate by
considering the instantaneous moment of inertia of the oblate star.  
Due to $J$-losses, $\lambda$ falls down, although the WD
may even rotate faster, leading to internal compression.

For the carbon-burning $^{12}$C+$^{12}$C reaction rate, we consider
the expressions given in \citet{1988ADNDT..40..283C}. A key ingredient
is the screening of this reaction. Here, we shall explore four different treatments, 
which are those of
\citet{1990ApJ...362..620I}, \citet{1997ApJ...481..883O},
\citet{2000ApJ...539..888K}, and \citet{2012A&A...538A.115P}.
\citet{2000ApJ...539..888K} computed the enhancement factors of
nuclear reaction rates due to interionic correlations within the
linear-mixing (LM) approximation, where polarization of electrons was
explicitly taken into account. More recently,
\citet{2012A&A...538A.115P} claimed that the deviation $\Delta h$ from
the LM approximation tends to reduce the enhancement factor
significantly, but their formula (19) is wrong; following their
notations, the correct formula should read

\begin{eqnarray}
 & \Delta h \equiv h_0-h_{\rm lm}= \nonumber \\
 & f_{\rm mix}(x_{\rm C}, x_{\rm O}, 0)
 -{d f_{\rm mix}(x_{\rm C}-(2-x_{\rm C})\xi,x_{\rm O}(1+\xi),\xi)
    \over d\xi}\biggr\vert_{\xi\to 0}, 
\end{eqnarray}

\noindent where $f_{\rm mix}(x_{\rm C}, x_{\rm O}, x_{\rm Mg})$ represents the
deviation of the interaction free energy from the LM value for C-O-Mg
three-component plasma, with $x_i$ ($i$=C, O, Mg) denoting their molar
fractions. When this correct formula is adopted, $\Delta h$ turns out
negligibly small under the present parameter conditions. For practical
use, they also suggested a simple Salpeter-Van Horn-type interpolation
formula combined with the quantum correction, but we shall not use it
here since it underestimates the reaction rates by one or two orders
of magnitude near the freezing point, exhibiting large discrepancies
with the pycnonuclear rates in the solid phase
\citep{2000ApJ...539..888K}.  We also note that
\citet{1990ApJ...362..620I} and \citet{1997ApJ...481..883O} ignored
the electron polarization effects. Consequently, we shall consider the
treatment by \citet{2000ApJ...539..888K} as the standard case and is
the one employed unless stated otherwise.

%-------------------------------------------------------------------------------
\section{RESULTS} \label{sec:resultados}

In this work, we restrict ourselves to the case of $\dot{M} =
10^{-6}\; M_{\odot}$ yr$^{-1}$; such material is burned in the
outermost WD layers to C+O composition. The initial composition is
of $X(^{12}{\rm C})= X(^{16}{\rm O}) = 0.49$ and heavier elements have
solar abundances.

We computed several evolutionary situations that depend on the binary
parameters (orbital period $P$ -- donor mass $M_2$) as shown in \S\ref{sec:discu_conclu}, 
stopping calculations
before the onset of hydrodynamical stages of thermonuclear runaway
(log$(T_{\rm c}/{\rm K})\sim 8.8$; \citealt{1982ApJ...257..780N}).

(1) One is the case in which the WD accretes material continuously up
to ``prompt'' carbon ignition. As described in \S\ref{sec:introd}, we obtained
$M_{\rm Ig}^{\rm R} =$ 1.43~$M_{\odot}$ for 
$\dot{M} = 10^{-7} M_{\odot}$ yr$^{-1}$.  This is smaller than the carbon
ignition mass of 1.48~$M_{\odot}$ obtained by \citet{2003ApJ...595.1094U} because
their 2D uniformly rotating model neglects the
Coulomb term in the equation of state.

\begin{figure} \begin{center}
  \includegraphics[scale=0.40,angle=0]{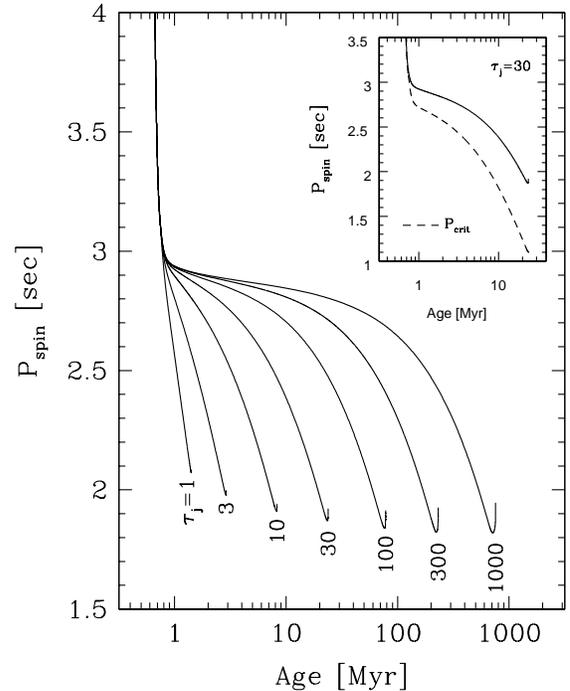}
\caption{\label{fig:spin} Spin period for the presupernova evolution of a WD
  for the case of screening of carbon-burning reaction given by
  \citet{2000ApJ...539..888K}. The final mass is of
  1.40~$M_{\odot}$. Accretion occurs for $P_{\rm spin}\gtrsim
  2.9$~sec. Each curve is labeled with its corresponding value of
  $\tau_{J}$ given in million years. Although the star loses
  angular momentum exponentially, its moment of inertia decreases fast
  enough to make the star to spin up after the end of
  accretion. However, remarkably, the star does not overflow its
  critical rotation rate avoiding shedding mass. This is shown in the
  inset for the case of $\tau_{J}=$~30~Myr, where spin and critical
  spin period evolution are given as solid and dashed lines,
  respectively.}
\end{center} \end{figure}

\begin{figure} \begin{center}
  \includegraphics[scale=0.330,angle=-90]{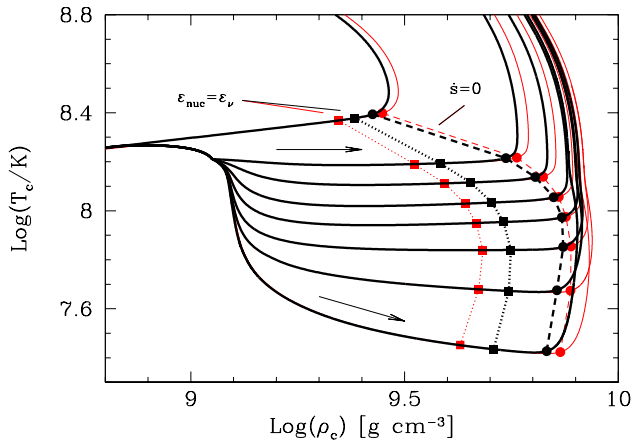}
\caption{\label{fig:all_central} Evolutionary tracks of the center
  of the WD up to the onset of the hydrodynamical
  explosion. The upper track corresponds to  ``prompt'' carbon ignition, 
  while the others are ordered from top to bottom with increasing 
  $J$-loss timescale ($\tau_{J}=$~1, 3, 10, 30, 100, 300, and 1000~Myr, respectively).
  $\varepsilon_{\rm nuc}=\varepsilon_{\nu}$ indicates the
  conditions at which neutrino losses equal nuclear energy release,
  while $\dot{S}=0$ shows the stages at which central entropy per
  baryon begins to increase. Thick black and thin red lines correspond
  to the treatments of screening given by \citet{2000ApJ...539..888K}
  and \citet{2012A&A...538A.115P}, respectively. Arrows indicate the
  sense of the evolution. (A color version of this figure is available
  in the online journal.).}
\end{center} \end{figure}

\begin{figure} \begin{center}
  \includegraphics[scale=0.30,angle=0]{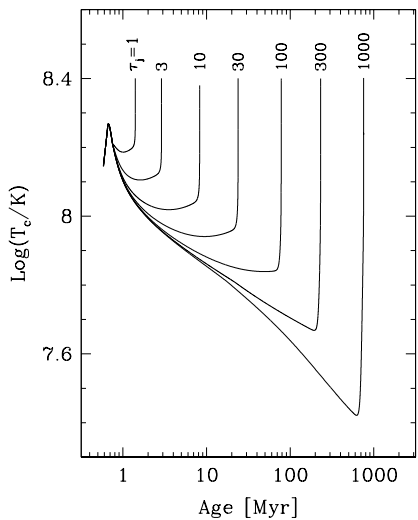}
\caption{Evolution of the central temperature as a function of
  time for the cases included in Figure~\ref{fig:spin}. \label{fig:tc_evol}}
\end{center} \end{figure}

\begin{figure} \begin{center}
  \includegraphics[scale=0.40,angle=0]{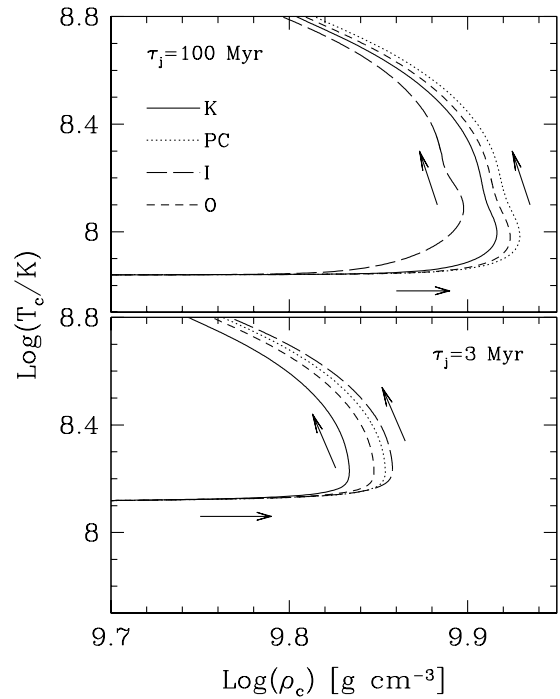}
\caption{\label{fig:final_central} Evolutionary tracks for the center
  of the WD during carbon ignition for the cases of $\tau_{J}=$~3 and
  100~Myr. I, O, K, and PC correspond to the results employing the
  treatments for the screening given in \citet{1990ApJ...362..620I},
  \citet{1997ApJ...481..883O}, \citet{2000ApJ...539..888K}, and
  \citet{2012A&A...538A.115P}, respectively. Arrows indicate the sense of the evolution.}
\end{center} \end{figure}

(2) We also consider a binary system in which the accretion rate
declines and the WD reaches a final mass $M$ before carbon
ignition; in this work, we shall consider the case of $M= 1.40\
M_{\odot}$.  For $t>t_{0}$ we apply $J$-losses as given
in Equation (4) with timescales of $\tau_{J}=$~1, 3, 10, 30, 100, 300,
and 1000~Myr. As quoted above, $J$-loss
processes are highly uncertain; thus, we employ a wide range for
$\tau_{J}$ to find the dependence of WD
evolution with this parameter. The main results are presented in
Figures~\ref{fig:spin}-\ref{fig:final_central}.

In Figure~\ref{fig:spin}, we show the evolution of the spin period
$P_{\rm spin}$ for the presupernova evolution of a WD. Accretion
occurs for $P_{\rm spin}\gtrsim 2.9$~s.  After the end of accretion,
the WD loses angular momentum exponentially (Equation (4)).  Despite
that, its moment of inertia decreases fast enough for the WD to {\sl
  spin up} rather than spin down. We thus do not call such an
evolution as ``spin down,'' rather just ``$J$-loss'' evolution. However,
remarkably, the WD does not overflow the critical rotation rate
avoiding shedding mass.  This is shown in Figure~\ref{fig:spin} for the
case of $\tau_{J}=$~30~Myr.  This is crucial for the delayed explosion
scenario to work because a little mass loss would be enough to preclude
the occurrence of the explosion.  In all cases, $P_{\rm spin}$
decreases (i.e., spin up rather than spin down) up to a minimum value
slightly before explosion; however, $\lambda$ decreases monotonically.

In Figure~\ref{fig:all_central}, we present the evolution of
the center of the WD since before the end of accretion up to
the onset of the hydrodynamical stage.  Thick black and thin red lines
correspond to the strong screening factors given by
\citet{2000ApJ...539..888K} and \citet{2012A&A...538A.115P},
respectively.  The uppermost line corresponds to the case (1)
evolution that leads to the ``prompt'' carbon ignition. Below that,
the lines from upper to lower correspond to the evolutions for
$\tau_{J}=$~1, 3, 10, 30, 100, 300, and 1000~Myr, respectively, that
lead to the ``delayed'' carbon ignition.

In Figure~\ref{fig:all_central}, we also show the 
conditions at which neutrino losses equals nuclear
energy release ($\varepsilon_{\rm nuc}= \varepsilon_{\nu}$)
and those that indicate when the central entropy per baryon begins to
increase $\dot{S}=0$, i.e., carbon ignition. These curves are well
detached because entropy per baryon starts to increases only when
nuclear energy releases exceeds neutrino and
conductive losses.  Conductive losses are comparatively more relevant
the longer the $\tau_{J}$ because neutrino losses are more
sensitive to temperature than conductive losses.

In Figure~\ref{fig:tc_evol}, we depict the evolution of the central
temperature as a function of time for the cases included in
Figure~\ref{fig:spin}.  It is seen that for longer $\tau_{J}$, the WD
is cooler and takes longer time until the carbon ignition occurs.

In Figure~\ref{fig:final_central}, we present the evolutionary tracks
for the center of the WD from the carbon ignition for the cases of
$\tau_{J}=$~3 and 100~Myr for the four strong screening factors (I, O,
K, and PC) given in \citet{1990ApJ...362..620I},
\citet{1997ApJ...481..883O}, \citet{2000ApJ...539..888K}, and
\citet{2012A&A...538A.115P}, respectively.  Note that
ignition densities are not so different among the four factors.

If $\tau_{J} \gtrsim 3$~Gyr, the central
temperature, and thus the rate of nuclear energy release, becomes so low
that carbon is depleted before reaching the
dynamical stage.  Due to electron capture, such a case could lead to collapse of the WD rather
than explosion, although a detailed study of this phenomenon is necessary (e.g.,
\citealt{2010ApJS..190..334F}).

%-------------------------------------------------------------------------------
\section{CONCLUSIONS AND DISCUSSION} \label{sec:discu_conclu}

The calculations presented above show that a C+O WD accreting enough
material coming from a binary companion, rotating uniformly at the
critical rate, and reaching a mass $M$ fulfilling $M_{\rm ig}^{\rm NR} <
M < M_{\rm ig}^{\rm R}$  (i.e., $1.38 M_\odot < M < 1.43 M_\odot$) 
may indeed undergo delayed explosion. The above
delay time is largely determined by the timescale of 
$J$-losses $\tau_{J}$ (see Figure~\ref{fig:tc_evol}). For the values of
$\tau_{J}$ considered here, the WD spends a time to undergo SN~Ia
explosion enough for the donor star to evolve to a structure
completely different from the one it had when acted as a donor.  For
the red giant donor, its H-rich envelope would be lost as a result of
H-shell burning and mass loss so that it would become a He WD in
$\sim$ 10 Myr.  For the main-sequence donor, it would also evolve to
become a low-mass He WD in 1 Myr, a hot He WD in 10 Myr, and a cold He
WD in 1000 Myr \citep{2011ApJ...738L...1D}. So, the 
$J$-losses should delay the explosion in enough time for the former donor
to be undetectable. Therefore, this scenario provides a way to account for
the failure in detecting companions to SNe~Ia.

\begin{figure} \begin{center}
  \includegraphics[scale=0.7,angle=0]{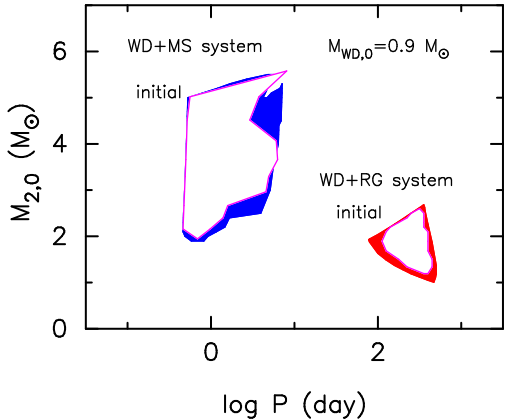}
  \caption{\label{fig:binary}
    Outcome of the binary evolution of the WD + companion star
    systems is shown in the parameter space of the initial orbital
    period $P$ and the companion mass $M_2$ for the initial WD mass
    of $0.9 M_\odot$.  The mass $M$ of the WD starting from the
    ``painted'' region reaches $1.38~M_\odot < M < 1.43~M_\odot$
    (delayed carbon ignition), while the systems starting from the
    blank region encircled by the solid line reach $M = 1.43~M_\odot$
    (prompt carbon ignition). }
\end{center} \end{figure}

Then, it is important to know for what binary systems the uniformly
rotating WD effectively stops increasing its mass $M$ at $1.38~M_\odot
< M < 1.43~M_\odot$ and undergoes the delayed carbon ignition.  We
have calculated the binary evolution of the WD + companion star
systems with the parameters of the initial orbital period $P$ and the
companion mass $M_2$ for the several initial WD masses as in
\citet{2012ApJ...756L...4H}.  The results for the initial WD mass of
$0.9 M_\odot$ is shown in Figure~\ref{fig:binary}.  Here, the binary
systems starting from the ``painted'' region of the ($P - M_2$) plane
reach $1.38~M_\odot < M < 1.43~M_\odot$, while the systems starting
from the blank region encircled by the solid line reach $M =
1.43~M_\odot$.  We estimate that the occurrence frequency of the
delayed carbon ignition would roughly be one-third of the total frequency of
the carbon ignition.  The occurrence frequency is smaller than the
estimate by \citet{2014MNRAS.445.2340W}, possibly because they adopted
the mass range of the $J$-loss evolution as $1.38~M_\odot < M <
1.5~M_\odot$.

While here we have restricted ourselves to present results
corresponding to the case of WDs that reach a mass of 1.40~$M_{\odot}$
with $\dot{M} = 10^{-6}\; M_{\odot}$ yr$^{-1}$, we have
explored this scenario for other final masses and accretion rates. We
found that the overall evolution of the objects is qualitatively the
same if the final WD mass is close enough to $M_{\rm ig}^{\rm R}$. While
we did not attempt to determine the lowest-mass value for ignition to
occur, we found that models with 1.37~$M_{\odot}$ do not ignite, while
those with 1.38~$M_{\odot}$ explode.
This is consistent with the adopted value of $M_{\rm Ig}^{\rm NR}$.

We explored different screening theories for the carbon-burning reaction finding that there is an uncertainty in the ignition density
of $\approx 10\%$. Remarkably, ignition (defined as the moment at
which entropy per baryon of matter at the center of the star begins to
increase) occurs at a density significantly higher than the one at
which nuclear energy release equals neutrino losses. This is due to
conductive losses, especially important for the case of large
$\tau_{J}$ values for which the WD interior is cool.

Carbon burning forms a convective core, which leads to the occurrence
of the convective URCA process (e.g., \citealt{2005MNRAS.356..131L}).
In the present study, we do not take into account this process,
assuming that it does not suppress the heating by carbon burning.
After the thermonuclear runaway, the central region would reach
nuclear statistical equilibrium (NSE) and undergo electron capture at
higher densities than in the non-rotating WD
\citep{1984ApJ...286..644N} as well as the prompt ignition.  This
would lead to the synthesis of more neutron-rich Fe-peak elements,
such as $^{54}$Fe, $^{56}$Fe, and $^{58}$Ni.  It would be important to
study such nucleosynthesis in multi-dimensional simulation of
convective deflagration, because the results would be sensitive to the
treatment of turbulent flame.

The nucleosynthesis results should be
compared with observational signatures of SNe Ia.  For example, the
emission lines observed in the spectra of the nebular phase of SNe Ia
indicate the existence of a large amount of stable Fe and Ni.
\citet{2007Sci...315..825M} estimated the amount of stable Fe is $\sim 0.2
M_\odot$ for most of normal SNe Ia (see also
\citealt{2015MNRAS.450.2631M} for SN 2011fe).  If such an amount of
stable Fe is mostly $^{54}$Fe and those SNe Ia are powered by the
radioactive decay of $\sim 0.6 M_\odot$ $^{56}$Ni, the ratio of
$^{54}$Fe/$^{56}$Fe in the ejecta of those SNe Ia is $\sim 8$ times
larger than the solar ratio.  This suggests that the stable Fe is
mostly $^{56}$Fe rather than $^{54}$Fe, which would be realized for lower $Y_e$, i.e., at
the higher central density than the prompt ignition model (such as W7
in \citealt{1984ApJ...286..644N}).  Lower $Y_e$ in the delayed carbon
burning might also be consistent with the large amount of Mn and (stable)
Ni are observed in the SN~Ia remnants
\citep{2015ApJ...801L..31Y}.

%-------------------------------------------------------------------------------
\acknowledgments

We would like to thank N. Itoh and S. Ogata for useful comments on the
strong screening factor.  This work has been supported in part by the
WPI Initiative, MEXT, Japan, and by the Grants-in-Aid for Scientific
Research of the JSPS (23224004, 24540227, and 26400222).

%-------------------------------------------------------------------------------

\end{document}